\documentclass[aps,prl,preprint,tightenlines,superscriptaddress,showpacs,byrevtex]{revtex4}
\newcommand{\blcpbpipi}{$\bar{B}^0\to\Lambda_c^+\bar{p}\pi^+\pi^-$}
\newcommand{\blclc}{$\bar{B}^0\to\Lambda_c^+\bar{\Lambda}_c^-$}

\newcommand{\bxilc}{$B^-\to\Xi_c^0\bar{\Lambda}_c^-$}
\newcommand{\xicxipi}{$\Xi_c^0\to\Xi^-\pi^+$}
\newcommand{\product}{${\cal B}$({\bxilc})$\cdot{\cal B}$({\xicxipi})}
\newcommand{\blcpb}{$\bar{B}^0\to\Lambda_c^+\bar{p}$}
\newcommand{\bscpb}{$B^-\to\Sigma_c^0\bar{p}$}

\newcommand{\upperbr}{${\cal B}(\bar{B}^0\to\Lambda_c^+\bar{\Lambda}_c^-)<6.2\times10^{-5}$} 
\newcommand{\brlclc} {$(2.2^{+2.2}_{-1.6}{\rm (stat)}\pm1.3{\rm (syst)})\times10^{-5}$}
\newcommand{\productbn}{${\cal B}(\bar{B}^0\to\Xi_c^+\bar{\Lambda}_c^-)\cdot{\cal B}(\Xi_c^+\to\Xi^-\pi^+\pi^+)=(9.3^{+3.7}_{-2.8}\pm1.9\pm2.4)\times10^{-5}$}
\newcommand{\ratioxi}{${\cal B}(\Xi_c^+\to\Xi^0\pi^+)/{\cal B}(\Xi_c^+\to\Xi^-\pi^+\pi^+)=0.55\pm0.16$} 
\newcommand {\brxicpls} {${\cal B}(\Xi_c^+\to\Xi_c^0\pi^+)=0.31-3.9\%$}
\newcommand {\brfourbody} {${\cal B}(\bar{B}^0\to\Lambda_c^+\bar{p}\pi^+\pi^-)=(1.12\pm0.05\pm0.14\pm0.29)\times10^{-3}$}
\newcommand {\brthreebody} {${\cal B}(B^-\to\Lambda_c^+\bar{p}\pi^-)=(2.01\pm0.15\pm0.20\pm0.52)\times10^{-4}$}
\newcommand {\brtwobody} {${\cal B}(\bar{B}^0\to\Lambda_c^+\bar{p})=(2.19^{+0.56}_{-0.49}\pm0.32\pm0.57)\times10^{-5}$}
\usepackage{graphicx} % Include figure files
\usepackage{dcolumn}  % Align table columns on decimal point

\graphicspath{{ps}}

\begin{document}

\preprint{\vbox{ \hbox{   }
%                 \hbox{Belle-Conference-0707}
%                 \hbox{pub-lclc.v3.12}
%                 \hbox{Belle Pub239}
                 \hbox{Belle Preprint 2007-41}
                 \hbox{KEK Preprint 2007-45}
}}

\title{  \quad\\[0.5cm] 
Search for {\blclc} decay at Belle
}
%\author{Y.Uchida, H.Ozaki and H.Kichimi}
%\author{Belle Collaboration}
%%% Paper:    
%%% Journal:  Summer 2007 Conference Papers
%%% July 17, 2007
%%% Contacts: 
%%% Non-responding authors or those who said NO are commented out.
%%% ====================================================================
%%% Click the RELOAD button on your web browser to see the updated file.
%%% ====================================================================
%%% Use \input{author} to insert this material into your latex file.
%%%%% Force institutions to appear in alphabetical order when typeset.
\affiliation{Budker Institute of Nuclear Physics, Novosibirsk}
\affiliation{Chiba University, Chiba}
\affiliation{University of Cincinnati, Cincinnati, Ohio 45221}
\affiliation{Department of Physics, Fu Jen Catholic University, Taipei}
\affiliation{Justus-Liebig-Universit\"at Gie\ss{}en, Gie\ss{}en}
\affiliation{The Graduate University for Advanced Studies, Hayama}
\affiliation{Gyeongsang National University, Chinju}
\affiliation{Hanyang University, Seoul}
\affiliation{University of Hawaii, Honolulu, Hawaii 96822}
\affiliation{High Energy Accelerator Research Organization (KEK), Tsukuba}
\affiliation{Hiroshima Institute of Technology, Hiroshima}
\affiliation{University of Illinois at Urbana-Champaign, Urbana, Illinois 61801}
\affiliation{Institute of High Energy Physics, Chinese Academy of Sciences, Beijing}
\affiliation{Institute of High Energy Physics, Vienna}
\affiliation{Institute of High Energy Physics, Protvino}
\affiliation{Institute for Theoretical and Experimental Physics, Moscow}
\affiliation{J. Stefan Institute, Ljubljana}
\affiliation{Kanagawa University, Yokohama}
\affiliation{Korea University, Seoul}
\affiliation{Kyoto University, Kyoto}
\affiliation{Kyungpook National University, Taegu}
\affiliation{Ecole Polyt\'ecnique F\'ed\'erale Lausanne, EPFL, Lausanne}
\affiliation{University of Ljubljana, Ljubljana}
\affiliation{University of Maribor, Maribor}
\affiliation{University of Melbourne, School of Physics, Victoria 3010}
\affiliation{Nagoya University, Nagoya}
\affiliation{Nara Women's University, Nara}
\affiliation{National Central University, Chung-li}
\affiliation{National United University, Miao Li}
\affiliation{Department of Physics, National Taiwan University, Taipei}
\affiliation{H. Niewodniczanski Institute of Nuclear Physics, Krakow}
\affiliation{Nippon Dental University, Niigata}
\affiliation{Niigata University, Niigata}
\affiliation{University of Nova Gorica, Nova Gorica}
\affiliation{Osaka City University, Osaka}
\affiliation{Osaka University, Osaka}
\affiliation{Panjab University, Chandigarh}
\affiliation{Peking University, Beijing}
\affiliation{University of Pittsburgh, Pittsburgh, Pennsylvania 15260}
\affiliation{Princeton University, Princeton, New Jersey 08544}
\affiliation{RIKEN BNL Research Center, Upton, New York 11973}
\affiliation{Saga University, Saga}
\affiliation{University of Science and Technology of China, Hefei}
\affiliation{Seoul National University, Seoul}
\affiliation{Shinshu University, Nagano}
\affiliation{Sungkyunkwan University, Suwon}
\affiliation{University of Sydney, Sydney, New South Wales}
\affiliation{Tata Institute of Fundamental Research, Mumbai}
\affiliation{Toho University, Funabashi}
\affiliation{Tohoku Gakuin University, Tagajo}
\affiliation{Tohoku University, Sendai}
\affiliation{Department of Physics, University of Tokyo, Tokyo}
\affiliation{Tokyo Institute of Technology, Tokyo}
\affiliation{Tokyo Metropolitan University, Tokyo}
\affiliation{Tokyo University of Agriculture and Technology, Tokyo}
\affiliation{Toyama National College of Maritime Technology, Toyama}
\affiliation{Virginia Polytechnic Institute and State University, Blacksburg, Virginia 24061}
\affiliation{Yonsei University, Seoul}
  \author{K.~Abe}\affiliation{High Energy Accelerator Research Organization (KEK), Tsukuba} % KEK
  \author{I.~Adachi}\affiliation{High Energy Accelerator Research Organization (KEK), Tsukuba} % KEK
  \author{H.~Aihara}\affiliation{Department of Physics, University of Tokyo, Tokyo} % Tokyo
  \author{K.~Arinstein}\affiliation{Budker Institute of Nuclear Physics, Novosibirsk} % BINP
  \author{T.~Aso}\affiliation{Toyama National College of Maritime Technology, Toyama} % Toyama
  \author{V.~Aulchenko}\affiliation{Budker Institute of Nuclear Physics, Novosibirsk} % BINP
  \author{T.~Aushev}\affiliation{Ecole Polyt\'ecnique F\'ed\'erale Lausanne, EPFL, Lausanne}\affiliation{Institute for Theoretical and Experimental Physics, Moscow} % ITEP
  \author{T.~Aziz}\affiliation{Tata Institute of Fundamental Research, Mumbai} % Tata
  \author{S.~Bahinipati}\affiliation{University of Cincinnati, Cincinnati, Ohio 45221} % Cincinnati
  \author{A.~M.~Bakich}\affiliation{University of Sydney, Sydney, New South Wales} % Sydney
% ...
% Lausanne
\collaboration{The Belle Collaboration}

\date{\today}

\begin{abstract}
We search for the doubly charmed baryonic decay {\blclc}, 
in a data sample of $520\times10^6$ $B\bar{B}$ events
accumulated at the $\Upsilon(4S)$ resonance with the Belle detector at
the KEKB asymmetric-energy $e^+e^-$ collider.  
We find no significant signal and set 
an upper limit of {\upperbr} at 90\% confidence level. 
The result is significantly 
below a naive extrapolation from ${\cal B}$({\bxilc})
assuming a simple Cabibbo-suppression factor of $|V_{cd}/V_{cs}|^2$.
The small branching fraction could be attributed to a suppression 
due to the large momentum of the baryonic decay products, which has been
observed in other charmed baryonic two-body $B$ decays.
\end{abstract}

%{\renewcommand{\thefootnote}{\fnsymbol{footnote}}

%% >>>>>> authorlist will go here

\maketitle

\normalsize

\newpage

\normalsize

\vskip 1.0cm
%%%%%%%%%%%%%

%\section{Introduction}
The large mass of the $b$ quark and the large quark mixing matrix element $V_{\mathrm cb}$~\cite{cabibbo,ckm} 
for the $b\to c$ transition lead to a large branching fraction ($\sim10\%$)~\cite{pdg2006} for charmed baryonic decays of the $B$ meson.
Charmed baryonic decays into four-, three- and two-body final 
states have already been observed. The measured branching fractions; 
{\brfourbody}~\cite{park}, {\brthreebody}~\cite{gaby4} and {\brtwobody}~\cite{gaby2}
(also see Ref.~\cite{gaby1,gaby3,cleo-lamc,cleo_blamc}), point to
a hierarchy of branching fractions depending on the multiplicity in the final state~\cite{beach04}.
The measurements provide stringent constraints on theoretical models for charmed baryonic decays of the $B$ 
meson~\cite{jarfi,chernyak,lamc2_last_theory}. 
 
\begin{figure}[!htb]
\centering
\includegraphics[width=0.30\textwidth]{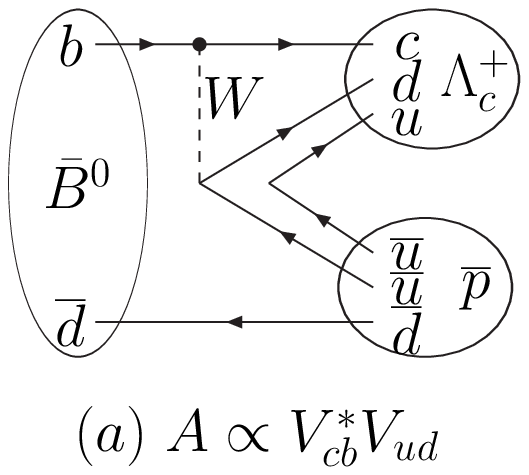}
\includegraphics[width=0.30\textwidth]{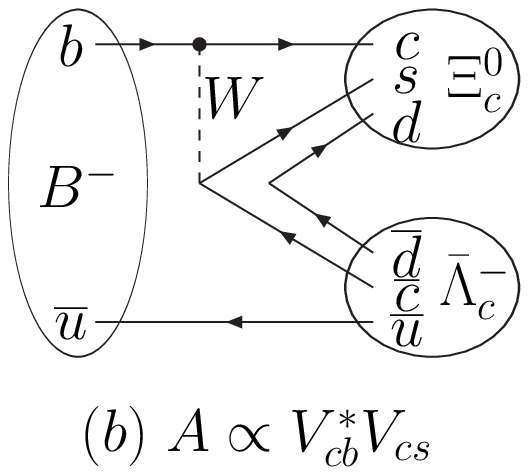}
\includegraphics[width=0.30\textwidth]{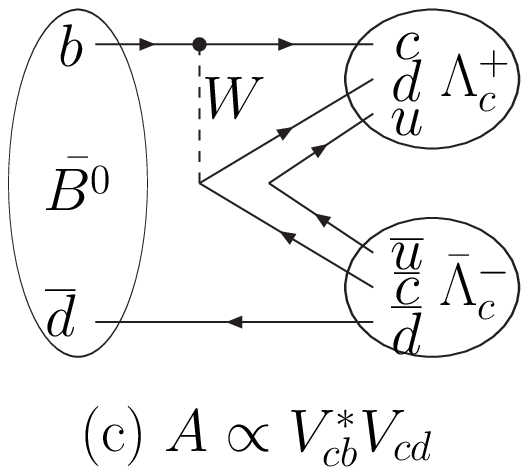}
\centering
\caption{
Quark diagrams for
(a) {\blcpb}, 
(b) {\bxilc} and
(c) {\blclc}.
The first two decays are Cabibbo-favored with CKM couplings 
$V_{cb}^*V_{ud}$ and $V_{cb}^*V_{cs}$, respectively,
while the third one is Cabibbo-suppressed with coupling $V_{cb}^*V_{cd}$. 
}
\label{2body-1}
\end{figure}

The hierarchy can be understood by large contributions of 
various intermediate states known in the decays~\cite{gaby1,park,gaby4,gaby2,cleo_blamc}.
The key is to understand quantitatively the decay mechanism of the two-body decays.
For example,
${\cal B}(B^-\to\Sigma_c(2455)^0\bar{p})=(3.7\pm0.7\pm0.4\pm1.0)\times10^{-5}$~\cite{gaby4}
is observed in the three-body decay $B^-\to\Lambda_c^+\bar{p}\pi^-$, which
is comparable to ${\cal B}$({\blcpb}).
There is also an interesting measurement of
${\cal B}$(\bxilc)=$(5.8\pm2.3)\times10^{-3}$~\cite{chistov,hycheng,qpxu,jgk,zencz},
which is quite large in comparison with ${\cal B}$({\blcpb}),
and does not follow the hierarchy.
Figures~\ref{2body-1}(a) and (b) show quark diagrams relevant for these decays 
through Cabibbo-favored $b\to c W^{-}$ transitions with
$W^-\to \bar{u}d$ and $W^-\to \bar{c}s$, respectively. 
Since we naively expect
similar branching fractions as $|V_{cb}^*V_{ud}|^2\sim|V_{cb}^*V_{cs}|^2$,
the two-order of magnitude difference between ${\cal B}$({\bxilc}) and ${\cal B}$({\blcpb}) 
is a puzzle.
It indicates that there is some mechanism to enhance or suppress specific two-body decays.
A discussion of a dynamical suppression mechanism,
based on the large Q-value in {\blcpb} compared to {\bxilc}, is given in Ref.~\cite{hychen}.
It is important to study various two-body decays 
to understand charmed baryonic $B$ decays.  

In this report, we study the doubly charmed baryonic decay {\blclc} 
as shown in Fig.~\ref{2body-1}(c). 
This mode is naively expected to have a branching fraction 
suppressed by a Cabibbo factor of $5.4\%$~\cite{pdg2006} relative to the Cabibbo-favored decays.
Given the large branching fraction ${\cal B}$({\bxilc}) relative to  ${\cal B}$({\blcpb}),
we search for the decay {\blclc}
and compare the observed branching fraction with simple estimates.
We expect ${\cal B}$({\blclc})=$(7.7\pm{3.0})\times10^{-7}$ 
from $\cal{B}$({\blcpb}),
taking into account the Cabibbo-suppression factor 
and the phase space factors in two-body decays proportional to the decay momentum
in the $B$ rest frame (assuming a relative S-wave, L=0).
Alternatively,
we expect ${\cal B}$({\blclc})=$(3.6\pm1.1)\times10^{-4}$
from ${\cal B}$(\bxilc).
We expect 0.1 and 45 events,  respectively,
from these two estimates scaled to our data sample

This analysis is based on a data sample of 479 fb$^{-1}$, corresponding to $520\times10^{6}$ $B\bar{B}$ events, 
which were recorded at the $\Upsilon(4S)$ resonance with the Belle detector at
the KEKB asymmetric-energy $e^+e^-$ collider~\cite{kekb}.

The Belle detector is a large-solid-angle spectrometer based on 
a 1.5~Tesla superconducting solenoid magnet. It consists of a three
layer silicon vertex detector for the first sample of $152\times10^6\,B\bar{B}$ pairs,
a four layer silicon vertex detector for the later $368\times10^6\,B\bar{B}$ pairs, 
a 50 layer central drift chamber (CDC), an array of aerogel threshold Cherenkov counters (ACC),
a barrel-like arrangement of time of flight scintillation counters (TOF), and an electromagnetic
calorimeter comprised of CsI(Tl) crystals located inside the superconducting solenoid coil.
An iron flux return located outside the coil is instrumented to detect $K_L^0$ mesons
and to identify muons. The detector is described in detail elsewhere~\cite{belle}. 
To simulate detector response and to estimate efficiency for signal measurement, 
we use Monte Carlo (MC) event generation program EvtGen~\cite{evtgen} and
a GEANT~\cite{geant} based detector simulation code. 

%\section{Event selection}
To search for {\blclc}
we reconstruct a pair of $\Lambda_c^+$'s decaying
into $p K^-\pi^+$.  Charge-conjugate modes 
are implicitly included throughout this paper unless noted otherwise.
We require tracks 
to have a distance of closest approach to the interaction point less 
than 5\,cm along the $z$-axis (opposite to the $e^+$ beam direction) and 1\,cm in a plane perpendicular to 
the $z$-axis.
Hadrons (protons, kaons and pions) are identified by using likelihood ratios based on  
CDC $dE/dx$, TOF and ACC information. 
We use likelihood ratios $L_{s}/(L_{s}+L_{b})$, where $s$ and
$b$ stand for the hadron species to be identified and for the others, 
respectively. We require the ratios to be greater than 0.6, 0.6
and 0.4 for proton, kaon and pion selection, respectively.
The efficiency for proton identification is 95\% with a kaon fake rate of 1.0\% due to 
the small proton momentum ($\sim1\,{\rm GeV}/c$) in these baryonic decays.
The efficiencies for kaons and pions are about 90\%,
while the corresponding pion and kaon misidentification rates are approximately 10\%~\cite{belle-pid}.
Tracks that are positively identified as electrons or muons are rejected.
We impose loose requirements on the vertex fit $\chi^2$'s for 
$\Lambda_{ {c}}^+\rightarrow p K^-\pi^+$ ($\chi^2_{\Lambda_c^+}$) and
{\blclc} ($\chi^2_{B}$)
to reject background from the decay products of $K_S^0$ 
and $\Lambda$ particles.
When there are multiple $B$ candidates (3\%) in an event,
we choose the candidate with the smallest $\chi^2_{B}$.

\begin{figure}[!htb]
\centering
\includegraphics[width=0.45\textwidth]{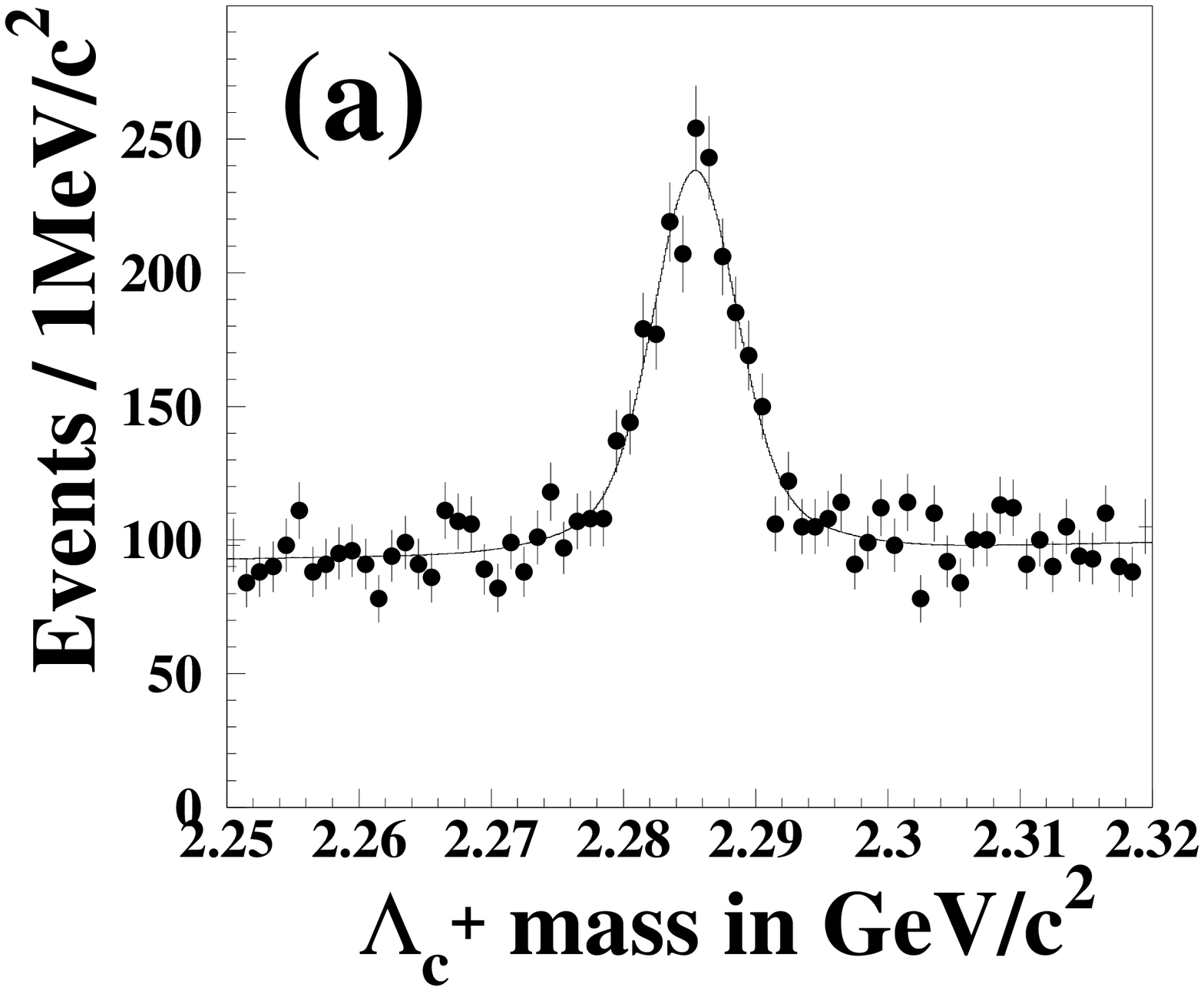}
\includegraphics[width=0.45\textwidth]{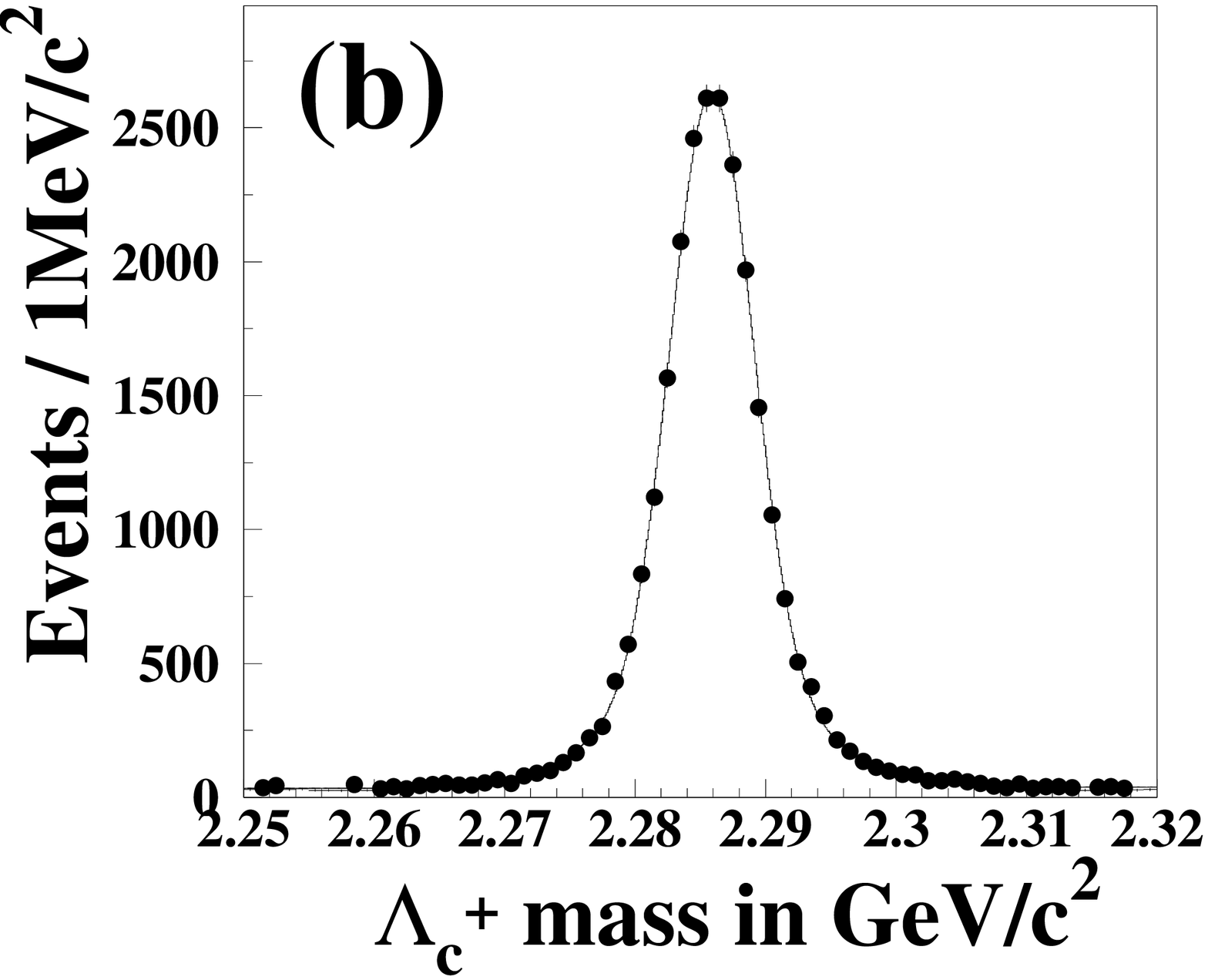}
\centering
\caption{
 $\Lambda_c^+(p K^- \pi^+)$ mass distribution for {\blclc} candidates
in $|\Delta{E}|<0.2\,{\rm GeV}$ and $5.2\,{\rm GeV}/c^2<M_{\rm bc}<5.3\,{\rm GeV}/c^2$.
 (a) Data and (b) MC signal.
 The curves show the fits with a double Gaussian for the signal and a linear function for the background.
}
\label{lamc-mass-fit}
\end{figure}

We search for the $B$ signal in the two dimensional plane of $\Delta{E}$ and $M_{\rm bc}$.
The variable $\Delta{E}=E_{B}-E_{\rm{beam}}$ is the difference between the reconstructed $B$ meson energy ($E_{B}$) and
the beam energy ($E_{\rm beam}$). $M_{\rm bc}=\sqrt{E_{\rm beam}^2 - P_B^2}$
is the beam energy constrained $B$ meson mass with
the momentum vector of the $B$ meson (${P_B}$).
Here $E_{\rm beam}$, $E_{B}$ and ${P_B}$ are defined in the center-of-mass system (CMS).
We use the $\Lambda_c^+$ mass~\cite{pdg2006} and the measured momentum of the $\Lambda_c^+$ system to calculate $E_B$, 
as it gives a better $\Delta{E}$ resolution, $4.3\,{\rm MeV}/c^2$, than that 
calculated with the $\Lambda_c^+$ energies reconstructed from the decay products, $6.6\,{\rm MeV}/c^2$. 
To optimize the selection parameters for the signal search,
we define a $B$ signal region of $|\Delta{E}|<0.02\,{\rm GeV}(\sim4\sigma)$ and 
$5.27\,{\rm GeV}/c^2<M_{\rm bc}<5.3\,{\rm GeV}/c^2$.

Figure~\ref{lamc-mass-fit} shows the $\Lambda_c^+$ mass distribution for (a) data and 
(b) the MC signal for $B$ signal candidates
with $|\Delta{E}|<0.2\,{\rm GeV}$ and
$5.2\,{\rm GeV}/c^2<M_{\rm bc}<5.3\,{\rm GeV}/c^2$.
We find a significant $\Lambda_c^+$ mass peak in the data due to the large inclusive branching fraction
for $B$ meson decays with a $\Lambda_c^+$ baryon in the final state.
The curves show fits using a double Gaussian for the signal and a linear function for the background.
We obtain a $\Lambda_c^{+}$ yield of $1281\pm69$ events with a $\chi^2/ndf=59.4/65$ (67.4\%).
In the fit to the data, we fix the ratio of $\sigma_{\rm tail}/\sigma_{\rm core}$ to 2.29 and the tail fraction 
(to the total area) to 0.284; these values are obtained from a fit to the MC signal.
The parameters $\sigma_{\rm tail}$ and $\sigma_{\rm core}$ are the widths for the core and tail Gaussians, respectively.
The fitted masses and $\sigma_{\rm core}$ are $(2285.3\pm0.2)\,{\rm MeV/c}^2$ and $(3.3\pm0.2)\,{\rm MeV/c}^2$
for the data, and $(2285.9\pm0.1)\,{\rm MeV/c}^2$ and $(3.2\pm0.1)\,{\rm MeV/c}^2$ for the MC signal.
We require that the $\Lambda_c^+$ masses lie in the range 2.275$\,{\rm GeV}/c^2$ to 2.295$\,{\rm GeV}/c^2$
($\pm3\sigma_{\rm core}$).
The small differences between the data and MC signal are taken into account in the systematic error 
as discussed later. 

In this analysis, the $\Lambda_c^+$ mass requirements are
very effective in suppressing the continuum background ($e^+e^-\to q {\bar q}, q=u,d,s,c$). 
The dominant background is from generic $B$ events. To suppress the background further, we use the variable
$\cos\theta_{\mathrm{B}}$, which is the cosine of the angle between the reconstructed $B$ direction and the $e^-$ 
beam direction in the CMS. 
The $B$ signal has a $(1-\cos^2\theta_{\mathrm{B}})$ distribution while the generic $B$ background 
and the continuum background have
a nearly flat distribution.
Using MC simulation, we examine the figure of merit $S/\sqrt{S+N}$ as a function
of $\cos\theta_{\mathrm B}$.
Here, $S$ and $N$ are the signal and background yields in the $B$ signal region, respectively. 
We assume a branching fraction ${\cal B}$(\blclc)$=5\times10^{-5}$ and a sample of $6\times10^{8}$ $B\bar{B}$ events,
and optimize the figure of merit with the requirement $|\cos\theta_{\mathrm B}|<0.8$.

%\section{Determination of Branching fraction}

To obtain the signal yield, we perform an unbinned 
maximum likelihood fit to the {\blclc} candidates in a two-dimensional (2D) region 
$-0.15\,{\rm GeV}<\Delta{E}<0.2\,{\rm GeV}$ and $5.2\,{\rm GeV}/c^2<M_{\rm bc}<5.3\,{\rm GeV}/c^2$.
We exclude the region $\Delta{E}<-0.15\,{\rm GeV}$, as we find from MC simulation
that a background from $B^{-/0}\to\Lambda_c^+\bar{\Lambda}_c^-\pi^{-/0}$
populates the region $\Delta{E}\sim-0.2\,{\rm GeV}$.
Thus, the effect of the background is
negligibly small ($<$0.05 events) in the fit region, even if we assume 
large values of ${\cal B}(B^{-/0}\to\Lambda_c^+\bar{\Lambda}_c^-K^{-/0})$~\cite{gaby3}.
%We also estimate a background of $0.3\pm0.2$ events from {\blcpbpipi}~\cite{park}, 
%when a  $\pi^{+}$ is misidentified as a $K^{+}$.}

We use a likelihood defined by
\begin{equation}
{L}={\frac { e^{ -(n_{\rm s}+n_{\rm b})} } {n!}}{\prod_{i=1}^{n}[n_{\rm s}F_{\rm s}(\Delta{E}_i,{M}_{{\rm bc}i})+n_{\rm b}F_{\rm b}(\Delta{E}_i,{M}_{{\rm bc}i})]}
\end{equation}  
with the signal yield $n_{\rm s}$ and the background yield $n_{\rm b}$. 
The parameter $n$ is the observed number of events.
The probability density function (PDF) for the signal $F_{\rm s}(\Delta{E},{M}_{{\rm bc}})$ is expressed as a product of a double Gaussian
in $\Delta{E}$ and a single Gaussian in $M_{\rm bc}$, while the PDF for the background
$F_{\rm b}(\Delta{E},{M}_{{\rm bc}})$ is expressed as a
product of a linear function in $\Delta{E}$ and an ARGUS function~\cite{argus_function} in $M_{\rm bc}$.

In the fit, the $\Delta{E}$ and $M_{\rm bc}$ signal shape parameters are fixed to those obtained from
one-dimensional fits to the individual simulated distributions for $\Delta{E}$ 
with $5.27\,{\rm GeV/c^2}<M_{\rm bc}<5.30\,{\rm GeV}/c^2$,
and $M_{\rm bc}$ with $|\Delta{E}|<0.02\,{\rm GeV}$.
The yields $n_{\rm s}$ and $n_{\rm b}$, the $\Delta{E}$ linear slope parameter and 
the ARGUS shape parameter are floated.
We obtain a signal efficiency of $0.106\pm0.001$ from a 2D fit to the MC signal.
For the fit to the data, we fix the signal parameters to those
calibrated for the MC/data systematic difference
by using a control sample of $\bar{B}^0\to\Lambda_c^+\bar{p}\pi^+\pi^-$ decays. 

\begin{figure}[!htb]
\centering
\includegraphics[width=0.90\textwidth]{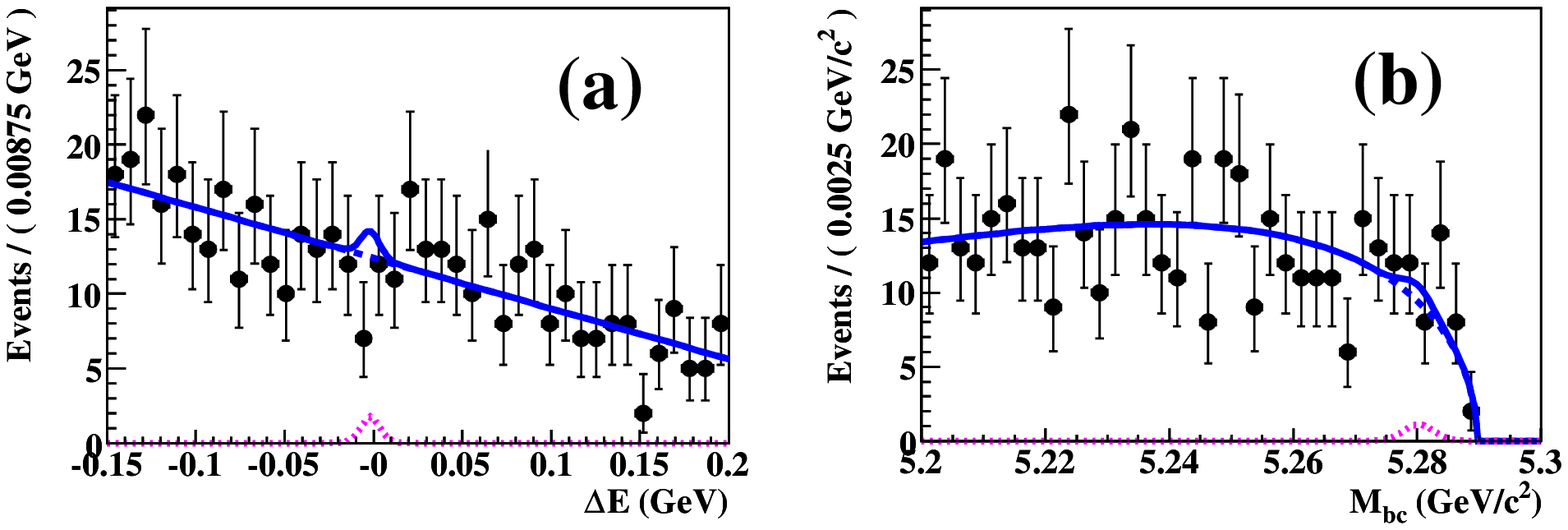}
\includegraphics[width=0.90\textwidth]{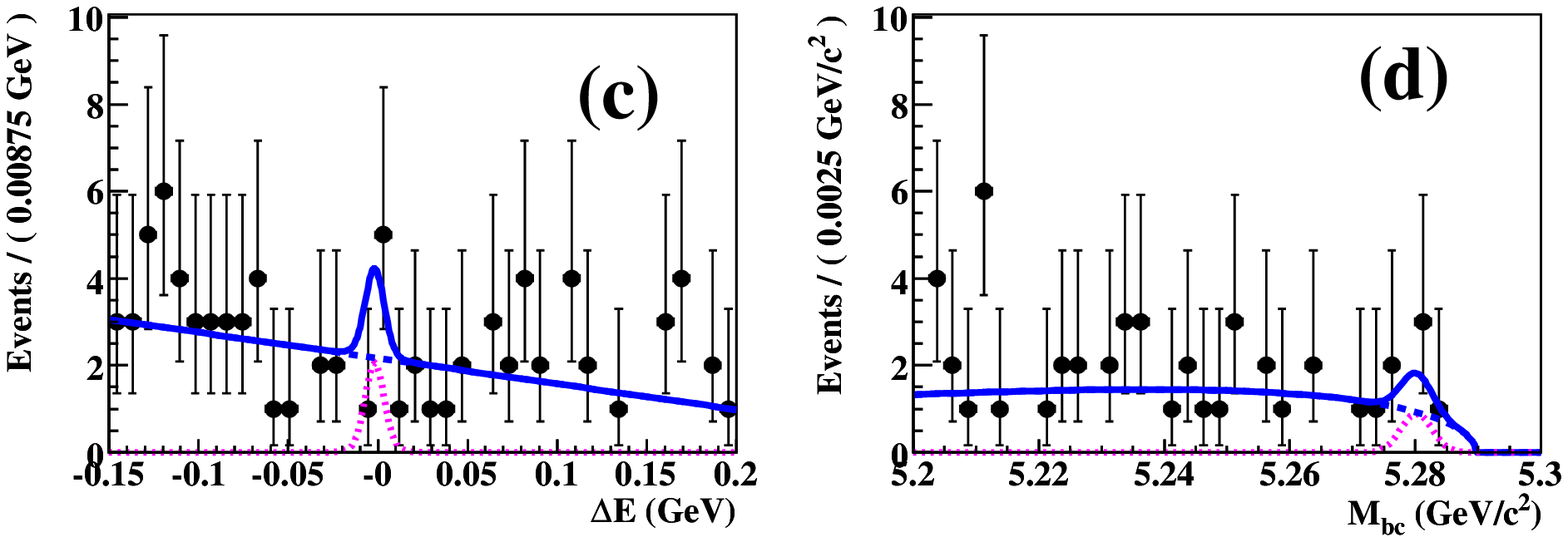}
\includegraphics[width=0.90\textwidth]{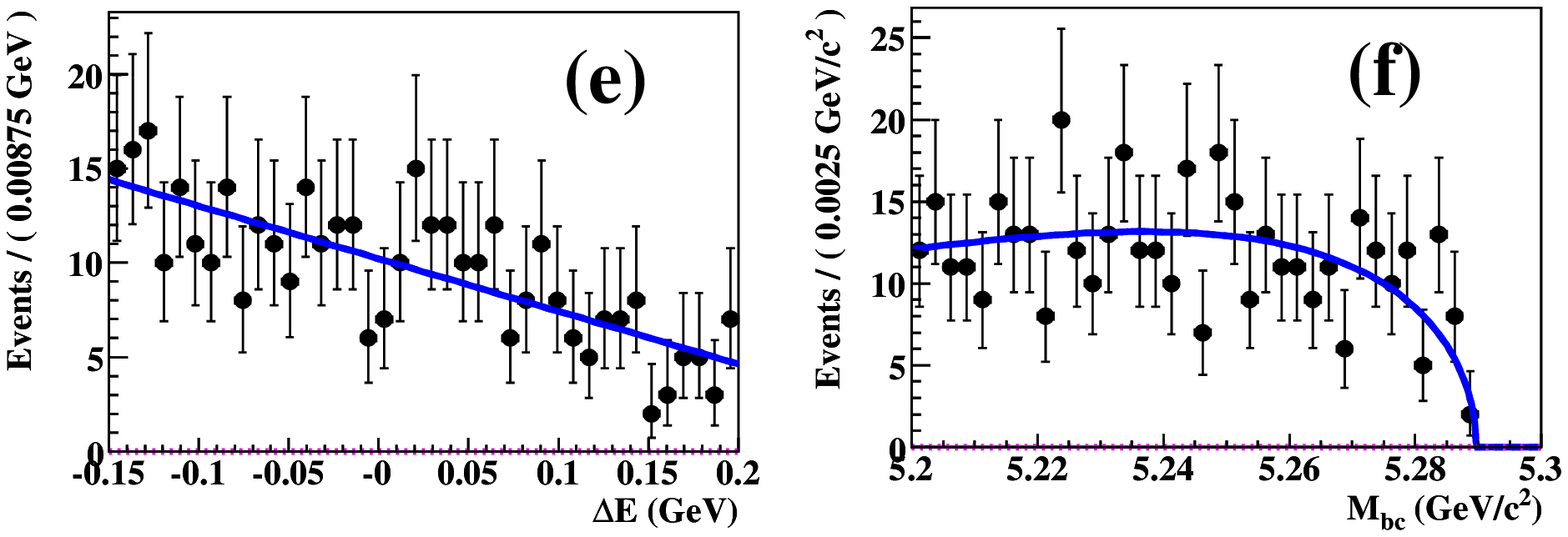}
\centering
\caption{
Two-dimensional unbinned likelihood fit to the data in $-0.15\,{\rm GeV}<\Delta{E}<0.20\,{\rm GeV}$
and $5.20\,{\rm GeV/c^2}<M_{\rm bc}<5.30\,{\rm GeV}/c^2$.
(a) $\Delta{E}$ and
(b) $M_{\rm bc}$ distributions for all events.
(c) $\Delta{E}$ distribution for $5.27\,{\rm GeV/c^2}<M_{\rm bc}<5.30\,{\rm GeV}/c^2$, and 
(d) $M_{\rm bc}$ distribution for $|\Delta{E}|<0.02\,{\rm GeV}$.
(e) $\Delta{E}$ distribution for $M_{\rm bc}<5.27\,{\rm GeV}/c^2$ and 
(f) $M_{\rm bc}$ distribution for $|\Delta{E}|>0.02\,{\rm GeV}$.
    The curves represent the fitted signal (dotted lines) and the total (solid lines) yield.
}
\label{2dfit-data}
\end{figure}

Figure~\ref{2dfit-data} shows the fit to the data.
We obtain a signal of $2.7^{+2.7}_{-2.0}$ events 
with a statistical significance of $1.6\sigma$. 
The significance is calculated as $\sqrt{-2{\rm ln}(L_0/L_{\rm max})}$, where
$L_{\rm max}$ and $L_0$ are the likelihood values at the fitted signal yield and the signal fixed to zero.

%\section{Background from \blcpbpipi ~and peaking background}
We investigate a possible peaking background in the sideband data, which includes a background
from {\blcpbpipi}~\cite{park}, when a  $\pi^{+}$ is misidentified as a $K^{+}$.
We define the sideband by requiring that one of the $\Lambda_c^+$ candidate masses
lies in the range $2.245\,{\rm GeV}/c^2-2.325\,{\rm GeV}/c^2$
while excluding masses in the range $2.275\,{\rm GeV}/c^2-2.295\,{\rm GeV}/c^2$.
From the 2D fit to the sideband, we estimate a peaking background of $-0.1\pm0.5$  
events, which is consistent with zero.
 
We estimate a systematic error of 14.5\% in event reconstruction and selection;
a 12.6\% uncertainty in the efficiency (arising from possible differences between the data and 
MC simulation in the reconstructed $\Lambda_c^+$ mass, particle identification and tracking),
a 7.1\% uncertainty due to the uncertainty of the signal parameterization used 
in the 2D fit (obtained by varying the parameters by one standard deviation),
and a 1.3\% uncertainty in the total number of $B{\bar B}$ events. 
We obtain a total systematic error of 62\% in the measured branching fraction,
including a 58\% uncertainty due to an error in ${\cal B}(\Lambda_c^+\to p K^- \pi^+)=(5.0\pm1.3)\%$~\cite{pdg2006}
and a 18\% error for the peaking background. 
We correct the signal efficiency by a factor of 0.90 due to a systematic difference
in particle identification between MC and data. 
We assume the same numbers of neutral and charged $B{\bar B}$ pairs, and obtain a branching fraction of {\brlclc}. 

We calculate 7.7 events for the upper limit yield at 90\% confidence level (CL) by integration of the 
likelihood function obtained from the 2D fit. We use the formula of
$ 90\% = { \int _{0}^{s_{\rm UL}} L(n|s) d{\mathrm s} }/{ \int _{0}^{\infty} L(n|s) d{\mathrm s} } $
with $n=2.7$, where 
the likelihood $L(n|s)= \int _{-\infty}^{\infty} L_{\rm fit}(n|s^*)\cdot G(s-s^*) d{\mathrm s^*}$ 
is convolved with the Gaussian $G(s-s^*)$ to take into account the total error,
which is composed of errors in the fitted yield (the signal and the peaking background), and
the systematic error discussed above.
The corresponding 
upper limit is found to be {\upperbr} at 90\% CL.

%\section{Discussion}

\begin{figure}[!htb]
\centering
\includegraphics[width=0.90\textwidth]{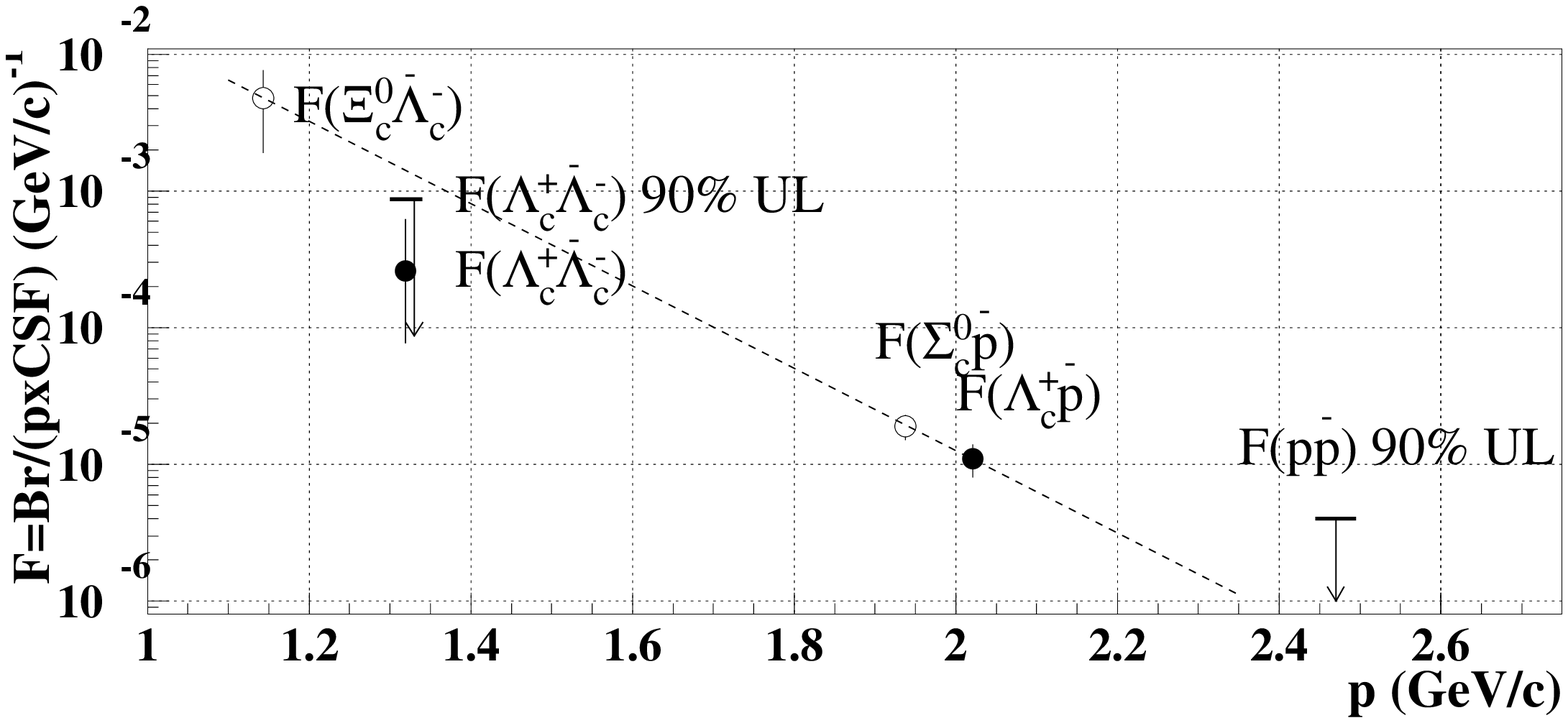}
\centering
\caption{
The rescaled branching fraction ${\cal F}={\cal B}/(p\cdot {\rm CSF})$ 
for {\bxilc}, {\blclc}, {\bscpb}, {\blcpb} and $\bar{B}^0\to p\bar{p}$ decays.
The dashed line shows a fit to ${\rm ln}({\cal F}(p))=c+s\times p$ 
with $s=-6.9\pm0.8\,({\rm GeV/}c)^{-1}$ to guide the eye.
}
\label{br-vs-qvalue}
\end{figure}

%%%%%BR.vs.Q PLOT ADDED
The present result is much smaller 
than the naive estimate of $(3.6\pm1.1)\times10^{-4}$ from ${\cal B}$({\bxilc})~\cite{chistov} with a significance
of approximately $3\sigma$, where the main uncertainty comes from the experimental error in {\product}.
On the other hand, 
no significant difference is observed 
for the naive estimate of $(7.7\pm{3.0})\times10^{-7}$ from Br(\blcpb)~\cite{gaby2}
due to the limited statistics.
Figure~\ref{br-vs-qvalue} compares the result 
with the data for other charmed baryonic two-body $B$ decays; {\bxilc}, 
$B^-\to\Sigma_c(2455)^0\bar{p}$~\cite{gaby4} and {\blcpb}.
We define a rescaled branching fraction ${\cal F}={\cal B}/({p}\cdot{\rm CSF})$~\cite{com1}.
Here $p$ is the decay momentum in the $B$ rest frame,
which represents a phase space factor (assuming L=0), and
CSF is a Cabibbo suppression factor~\cite{pdg2006}: 1.0 for {\bxilc} and {\blcpb}, 
and 0.054 for {\blclc}. 
We also plot ${\cal F}(p \bar{p})_{UL}$ for the 90\% CL upper limit on
${\cal B}(\bar{B}^0\to p\bar{p})$~\cite{tsai} with
CSF=$|V_{ub}/V_{cb}|^2=0.011$~\cite{pdg2006}
assuming a $b\to u(d\bar{u})$ tree transition.
The open and solid points with error bars show the data for $B^-$ and $\bar{B}^0$ decays, respectively.
The dashed line shows the function ${{\rm ln}({\cal F}(p))}=c+s\times p$ with $s=-6.9\pm0.8\,({\rm GeV/}c)^{-1}$ 
to guide the eye, which is obtained by a fit to the three data points. 
The 90\% CL upper limit ${\cal F}(\Lambda_c^+\bar{\Lambda}_c^-)_{UL}$ is
close to the line. 

In summary,
we search for the doubly charmed baryonic decay {\blclc} 
in a data sample of $520\times10^{6}$ $B\bar{B}$ events.
We obtain ${\cal B}$({\blclc})={\brlclc} with
an upper limit of {\upperbr} at 90\% confidence level.
The result is significantly smaller 
than a naive extrapolation 
from ${\cal B}$(\bxilc), assuming a simple Cabibbo suppression factor.
The suppression of {\blclc} could be attributed to 
the strong momentum dependence of the decay amplitude that has been 
observed in other charmed baryonic two-body $B$ decays.
 
\vspace {0.5cm}
%ACKNOWLEDGMENT
%\vspace {0.5cm}

%-------- Short version, if necessary, for PRL -----------
We thank the KEKB group for excellent operation of the
accelerator, the KEK cryogenics group for efficient solenoid
operations, and the KEK computer group and
the NII for valuable computing and Super-SINET network
support.  We acknowledge support from MEXT and JSPS (Japan);
ARC and DEST (Australia); NSFC and KIP of CAS (China); 
DST (India); MOEHRD, KOSEF and KRF (Korea); 
KBN (Poland); MES and RFAAE (Russia); ARRS (Slovenia); SNSF (Switzerland); 
NSC and MOE (Taiwan); and DOE (USA).


\begin{thebibliography}{99}
\bibitem{com1}
For comparison of those data points, 
we normalize them to the absolute branching fraction ${\cal B}(\Lambda_c^+\to p K^- \pi^+)=5.0\%$.
We calculate the errors for ${\cal F}(\Xi_c^0\bar{\Lambda}_c^-)$, ${\cal F}(\Sigma_c^0\bar{p})$, ${\cal F}(\Lambda_c^+\bar{p})$
excluding the uncertainty from the $\Lambda_c^+$ absolute branching fraction as these modes decay to final states
with one $\Lambda_c^+$. On the other hand, 
we include the error for ${\cal F}$(\blclc) as it decays to two $\Lambda_c$'s. 
The error in ${\cal F}(\Xi_c^0\bar{\Lambda}_c^-)$ includes the theoretical uncertainty in
${\cal B}$({\xicxipi})=0.83-1.74\% discussed above.
\end{thebibliography}
\end{document}